# MoS$_2$ Field-effect Transistors with Graphene/Metal Heterocontacts

Yuchen Du, Lingming Yang, Jingyun Zhang, Han Liu, Kausik Majumdar, Paul D. Kirsch, and Peide D. Ye, *Fellow, IEEE*

*Abstract*— For the first time, n-type few-layer MoS$_2$ field-effect transistors with graphene/Ti as the hetero-contacts have been fabricated, showing more than 160 mA/mm drain current at 1 µm gate length with an on-off current ratio of 10$^7$. The enhanced electrical characteristic is confirmed in a nearly 2.1 times improvement in on-resistance and a 3.3 times improvement in contact resistance with hetero-contacts compared to the MoS$_2$ FETs without graphene contact layer. Temperature dependent study on MoS$_2$/graphene hetero-contacts has been also performed, still unveiling its Schottky contact nature. Transfer length method and a devised I-V method have been introduced to study the contact resistance and Schottky barrier height in MoS$_2$/graphene /metal hetero-contacts structure.

*Index Terms*—MoS$_2$, Graphene, Hetero-contacts, MOSFET, Schottky barrier height

## I. INTRODUCTION

GRAPHENE and other two-dimensional (2D) materials are the rapidly rising stars on the horizon of materials science, condensed-matter physics, and solid state devices. They stand for a brand new family of materials that are one or few atomic layer thick, and offer many new research directions towards nanoscience and nanotechnology. However, the gapless nature of monolayer graphene has restrained its wide electronic device applications, in particularly for logic circuits [1]. Transition metal dichalcogenides (TMDs), another type of 2D materials, recently attracted wide attentions due to their appropriate bandgap and reasonable mobility. Owing to its unique structural and electronic properties, TMDs provide us novel material systems to explore interesting phenomena, such as 2D interfaces, which might be not available to be investigated in the traditional Si and III-V semiconductors. As one of the most studied TMDs, MoS$_2$ has a bandgap between 1.3-1.8 eV, depending on the number of layers, and reasonable mobility [2,3]. The on-state performance of the MoS$_2$ FETs is mainly limited by the large contact resistance at MoS$_2$/metal interfaces [4-9]. Recent attempts to improve MoS$_2$ contacts had been concentrated on the following areas using:(1) low work function contact metal [10], (2) gas doping of MoS$_2$ flakes [11], and (3) molecular or solid doping on MoS$_2$ films [12,13]. However, the fundamental reason for the large contact resistance is the result of Fermi level pinning on MoS$_2$ near conduction band edge due to S-vacancy defect level and charge neutral level location [14,15]. All above experiments have just partially touched the issue and barely resolved the issue completely. Graphene contacts on MoS$_2$ [9], which has a 2D to 2D interface, might provide a new angle to solve this technical challenge. In this letter, graphene is used in between metal contact and n-type few-layer MoS$_2$ film to enhance the electronic coupling between metal and MoS$_2$ and boost the electron injection into MoS$_2$. For the first time, well-performed graphene/metal hetero-contacts MoS$_2$ FETs were fabricated. The feasibility using graphene/ metal hetero-contacts to reduce contact resistance ($R_c$) and improve on-resistance ($R_{on}$) of the devices is demonstrated.

## II. EXPERIMENT

MoS$_2$ flakes with 5-6nm thickness were mechanically exfoliated from bulk ingot (SI Supplies) by standard scotch tape technique, and then transferred to a heavily p-doped silicon substrate with a 90 nm SiO$_2$ capping layer. Monolayer CVD graphene grown on Cu foil (Graphene Supermarket Inc.) was transferred onto the MoS$_2$/SiO$_2$/Si wafer with the PMMA method: A 500 nm PMMA (10% in Anisole) layer was first spin-coated on graphene/Cu. The copper substrate was etched using 1M FeCl$_3$ solution. After rinsing in DI water for several times, the PMMA/graphene layer was transferred onto the target substrate. The sample was dried in N$_2$ ambient for 12 hours, and then PMMA was removed by rinsing in acetone bath. An additional H$_2$/N$_2$ annealing at 450 $^o$C was used to remove the PMMA residues. After the graphene film transfer onto the MoS$_2$/SiO$_2$/Si substrate, electron beam lithography was used to pattern the source and drain contacts, combining with low power oxygen etch to isolate the source/drain, and the different devices. Graphene is fully etched from the channel. Metallization was performed by electron-beam evaporation of 20 nm/60 nm Ti/Au with the size 50 nm smaller at each side than graphene contacts as illustrated in Figure 1. In order to have accurate comparison, similar device without monolayer graphene were also fabricated on few-layer MoS$_2$ with a similar thickness and the same channel length. Electrical measurements were carried out with Keithley 4200 semiconductor parameter analyzer and probe station in ambient atmosphere.

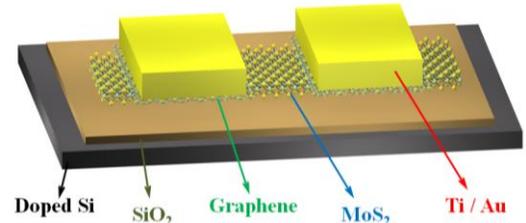

Figure 1. Structure geometry of hetero-contacts MoS$_2$/Graphene/Metal FETs with a channel length of 1 µm.

## III. RESULTS AND DISCUSSION

Output and transfer curves of a hetero-contacts FET are plotted in Figure 2(a) and (b) to demonstrate its effectiveness to improve device performance by inserting monolayer graphene between metal and MoS$_2$ contacts. The output curves of a

Yuchen Du, Lingming Yang, Jingyun Zhang, Han Liu, and Peide D. Ye are with the School of Electrical and Computer Engineering and Birck Nanotechnology Center, Purdue University, West Lafayette, IN 47907 USA (e-mail: yep@purdue.edu).

Kausik Majumdar and Paul D. Kirsch are with SEMATECH, 257 Fuller Road, Albany, NY 12203, USA

The work is partly supported by SEMATECH and SRC under Tasks 2362 and 2396.

reference sample are also shown in Figure 2(a) in black. For the device contact without graphene, shown in Figure 2(a),

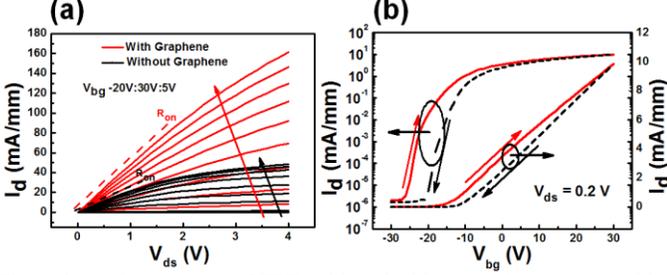

Figure 2. (a) Output curve of FETs with and without graphene contacts. The backgate voltage sweeps from -20 V to 30 V with a step of 5 V. (b) Transfer curve of the hetero-contacts FETs in log (left) and linear (right) scale. Both forward (solid) and backward (dash) sweeps are shown.

on-current at $V_{ds}$=4 V, and $V_{bg}$ = 30 V is equal to 48.5 mA/mm. As the monolayer graphene applied in the hetero-contacts, on-current is improved to be 161.2 mA/mm at the same source/drain and gate bias. $R_{on}$ determined from linear region as shown in Figure 2(a) as dashed lines are used to have a direct comparison. Comparing to $R_{on}$ of 42.8 Ω·mm for the reference device, hetero-contacts device shows a much lower $R_{on}$ of 20.6 Ω·mm. In order to have a more accurate comparison, metal contacts with and without graphene on the same flake have also been fabricated. Similar values of $R_{on}$ and similar amount of reduction in $R_{on}$ have been achieved. The results further verified the feasibility of using graphene/metal hetero-contacts to improve on-resistance of the devices. With similar channel resistance of two devices, the nearly 2.1 times dropping in $R_{on}$ is directly related to the reduction of $R_c$ by inserting monolayer of graphene into contact. In Figure 2(b), the transfer curves of hetero-contacts FETs are plotted with source/drain biases of 0.2 V. The exhibiting hysteresis from forward and backward sweeps is due to the charge injection at the interface between $MoS_2$ and the substrate and the charge transfer from/to neighboring adsorbates [16]. The $10^7$ current on/off ratio is achieved on hetero-contacts devices. Extrinsic field-effect mobility of hetero-contacts FETs is calculated to be 32.3 $cm^2/Vs$, which is extracted from the forward I-V transfer curves by the relation,

$$\mu = \frac{d\,I_{ds}}{d\,(V_{bg})} \times \frac{L}{W \times C_{ox} \times V_{ds}} \quad (1)$$

where $\frac{d\,I_{ds}}{d\,(V_{bg})}$ is the transconductance, L is the channel length, W is the channel width, and $C_{ox}$ is the backgate capacitance per unit area. Using $R_c$ of hetero-contacts device ~3.7 Ω·mm, intrinsic field-effect mobility of 50.4 $cm^2/Vs$ is obtained.

A devised I-V method has been used to quantitatively investigate the Schottky barrier height of hetero-contacts of $MoS_2$/graphene/metal [17]. The barrier height is calculated from the current $I_s$, determined by extrapolating the semilog $I_d$ versus $V_d$ curve to $V_d$=0 V. Schottky barrier height $\phi_B$ is calculated from $I_s$ according to

$$\phi_B = \frac{K_B T}{q} \ln(\frac{A \times A^* \times T^2}{I_S}) \quad (2)$$

where A is the area, $A^*$ is the Richardson constant, and T is the temperature. Due to the uncertainty of $A^*$ [17], the effective Schottky barrier height cannot be accurately calculated directly, however, instead of calculating the exact value of Schottky barrier height, exponential function of barrier height can be readdressed. Normalized exponential Schottky barrier height versus backgate voltage has been plotted in Figure 3(a) to demonstrate the Schottky barrier deduction with graphene contact. Moreover, transfer length method (TLM) with separations of 0.5 μm, 1 μm, 1.5 μm, and 2 μm are used to accurately extract the contact resistance of $MoS_2$/graphene/metal hetero-contacts. Contact resistance without graphene has also been extracted by applying TLM structure on a similar thickness flake. In Figure 3, both Schottky barrier and contact resistance show a strong gate bias dependent behavior. It is because (1) gate-tunable Schottky barrier width reduction due to the increment of backgate bias and (2) carrier concentration enhancement from backgate bias doping, where all lead to reduce the effective Schottky barrier height and generate more injected electrons from Fermi level at metal to conduction band of $MoS_2$. To exclude the large absolute errors, high bias regions are appropriate to have direct comparison in the contact resistance. Contact resistance with monolayer graphene is 3.7±0.3 Ω·mm at $V_{bg}$ = 30 V, which decreased from a value of 12.1±1.2 Ω·mm at the same backgate voltage without the graphene layer. This nearly 3.3 times reduction in contact resistance may possibly be attributed to the gate-induced electron injection from graphene layer to $MoS_2$. At the hetero-contact structure, the positive back-gate bias not only electro-statically dopes $MoS_2$, but also could move the Fermi-level in Ti doped n-type graphene [18-21] further up beyond the Ti/$MoS_2$ pinning level, thus enhance the electron injection from metal into the conduction band of $MoS_2$ leading to a lower contact resistance. The key difference from previous devices is that the back-gate can modulate not only the Fermi-level of the channel but also the contact (graphene) as shown in the inset of Figure 3(b) [19]. The total carrier density summed over graphene and $MoS_2$ hetero-contact exceeds the single $MoS_2$/metal contact, where the monolayer graphene can be seemed as a "charge pumping" layer. It seems that 2D graphene to $MoS_2$ interface is fundamentally different from metals to $MoS_2$ interfaces.

Temperature dependent measurements of graphene/ metal contact on $MoS_2$ have been also performed. Substrate temperatures are changed from 300 K to 400 K, and the electrical transport properties are reported in Figure 4. As the temperature goes up, thermal-assistant tunneling improves the electron injection efficiency from the hetero-contacts to $MoS_2$ channel through the Schottky barrier, thus lowering the contact resistance. Meanwhile, higher temperature also causes off-current $I_{off}$ to increase rapidly, where insulating behavior in off-state well fits to the activated temperature-dependent current. Calculated activation energy of $MoS_2$, $E_a$=0.23 eV, is the slope of Arrhenius plot in Figure 4(a). The thermal activity involved off-state conducting has the energy that is smaller than the band gap of $MoS_2$, which may correspond to the depth of the donor levels [22,23] most likely caused by natural material defects [14,15]. As depicted in Figure 4(c), the current on/off ratio drops with the increase of temperature, because of the increment of off-state current. Intrinsic field-effect mobility has been extracted in Figure 4(b). In the range of 300 K to 400 K, the temperature dependent intrinsic field-effect mobility is decreased from 50.4 $cm^2/Vs$ at 300 K to the lowest value of 28.7 $cm^2/Vs$ at 400K due to the electron-phonon scattering. Temperature dependence of the mobility follows the equation

$\mu \sim T^{-\nu}$, where the exponent ν depends on the dominating phonon scattering mechanism. From the generic temperature dependence fitting curve, we find the value of ν equals 1.87, which is larger than the value of 1.4-1.6 obtained from MoS$_2$ temperature dependent Hall mobility measurement [23,24]. The discrepancy could be related to the behavior of traps which affects field-effect mobility much more [25,26].

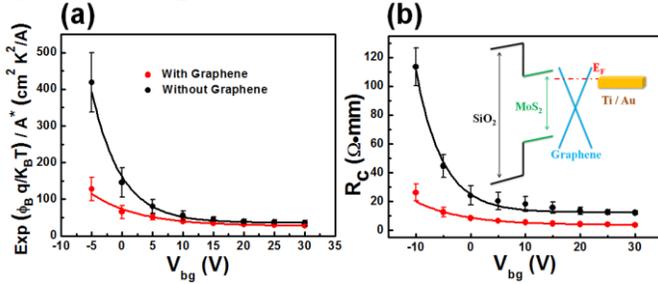

Figure 3. (a) Normalized exponential Schottky barrier height versus back-gate voltage (b) Contact resistance versus back-gate voltage for both contact with graphene and contact without graphene. Inset: Schematic band diagram of a metal/graphene/MoS$_2$ hetero-contact at a very positive gate bias with a zero drain bias. Error bars are determined from the standard errors of the linear fitting under different back-gate biases.

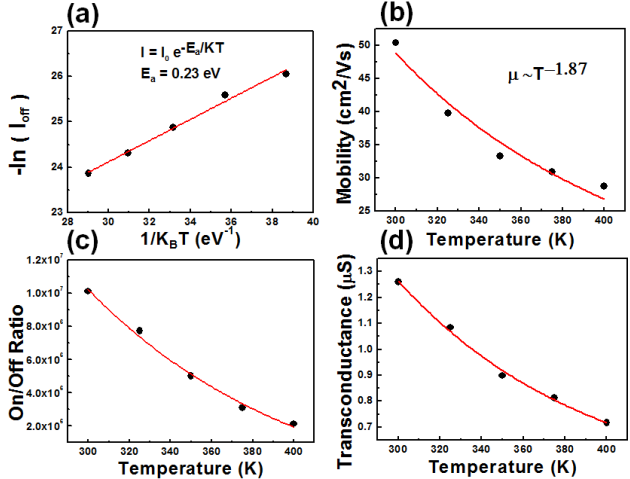

Figure 4. Temperature dependent measurement from 300 K to 400 K on (a) Off state current, (b) Intrinsic mobility, (c) On/off ratio, (d) Transconductance.

## IV. CONCLUSION

In summary, we have demonstrated the first MoS$_2$ field-effect transistor with graphene/metal hetero-contacts. Graphene hetero-contacts effectively reduce the contact resistance from 12.1 ±1.2 Ω·mm to 3.7 ±0.3 Ω·mm, compared to the reference MoS$_2$/metal contacts. The nearly 3.3 times improvement in R$_c$ is attributed to the gate-enhanced electron injection from metal doped n-type graphene into MoS$_2$ conduction band, which provides a new route to engineer 2D contacts towards the realization of Ohmic contacts on MoS$_2$ and other 2D materials.